\documentclass{ws-procs9x6}

\usepackage{latexsym}

\usepackage{epsfig,color}
\usepackage{epstopdf}
\newcommand{\be}{\begin{eqnarray}}
\newcommand{\ee}{\end{eqnarray}}

\usepackage{natbib}
\begin{document}

\title{Issues with Core-Collapse Supernova Progenitor Models}

\author{S.~W. Bruenn \footnote{\uppercase{W}ork partially
supported by grant from the \uppercase{DOE} \uppercase{O}ffice of  \uppercase{S}cience
\uppercase{S}cientific \uppercase{D}iscovery through \uppercase{A}dvanced \uppercase{C}omputing \uppercase{P}rogram.}}

\address{Florida Atlantic University, \\
777 West Glades Road, \\ 
Boca Raton, FL 33431, USA\\ 
E-mail: bruenn@fau.edu}

\maketitle

\abstracts{
The status of core collapse supernoova progenitor models is reviewed with a focus on some of the current uncertainties arising from the difficulties of modeling important macrophysics and microphysics. In particular, I look at issues concerned with modeling convection, the implications of the still uncertain $^{12}$C($\alpha,\gamma$)$^{16}$O reaction rate, the uncertainties involved with the incorporation of mass loss, rotation, and magnetic fields in the stellar models, and the possible generation of global instabilities in stellar models at the late evolutionary stages.  
}

\section{Inroduction}
\label{sec: Introduction}

The core-collapse supernova mechanism is still an unsolved problem. The failure of ``state-of-the-art'' one-dimensional core collapse simulations, utilizing multienergy-multiangle  neutrino transport schemes and realistic opacities, and the plethora of evidence that the phenomenon is inherently multidimensional has inaugurated a new era of supernova modeling.  Supernova codes are now beginning to couple multidimensional hydrodynamics to multidimensional neutrino transport, or at least neutrino transport along radial rays. Furthermore, spectacular advances are being made in the microphysics, particularly in the equation of state and the neutrino opacities. Many of the issues and prospect of both the microphysics and macrophysics involved in realistic core collapse supernova modeling are discussed in this volume.
 
Here I discuss the input data to these core collapse simulations, namely the progenitor models. Supernova modelers (myself included) tend to take delivery of these progenitor models without being fully cognizant of the approximations that are made in order to evolve these models from the main sequence to the point of core-collapse. Phenomena such as convection, rotational instabilities, and mass loss involve huge ranges of spatial and temporal scales and/or uncertain physics and require some sort of parameterization. Some of the macrophysics is inherently multidimensional. The purpose here is gather together most of the salient approximations and parameterizations that are made in computing progenitor models, and to increase thereby the awareness of the supernova community to the many aspects of current progenitor models open to future revision. Unfortunately, without a viable model of the core-collapse supernova mechanism, it is difficult to assess the effect on the core-collapse supernova scenario of variations within the uncertainties of the progenitor models. Conventional wisdom would point to the mass of the precollapse iron core, the density profile in the adjacent silicon and oxygen layers,  and its rate of rotation as being particularly important. The reader will find many details of recent progenitor models reviewed by \citet{maeder_m00a} and \citet{woosleyhw02}.

\section{Convection}
\label{sec: Convection}

Convection at the high Reynolds number characterizing flows inside stars is highly turbulent and chaotic characterized by eddies on a vast spectrum of scales. A great source of uncertainty in current stellar evolutionary calculations is how to model the thermal and compositional mixing at convective boundaries, and how to model the reactive flows that occur during late evolutionary phases when convective and nuclear time scales become comparable. First-principled numerical simulations of turbulence with the necessary resolution are not yet practicable, and needless to say it has been impossible to couple a first-principled calculation of turbulence with a stellar evolution code. Almost all stellar evolution codes model convection with some variant of ``mixing length theory'' (MLT) \citep{bohm_vitense_58} which attempts to capture the effects of convection by an essentially one parameter diffusion process. The convective diffusivity is taken to be $K_{\rm conv} = \frac{1}{3} v_{\rm conv} \ell$ where $v_{\rm conv}$ is the mean velocity of a typical convective eddy as it traverses a mean free path, $\ell$. The mean velocity is computed from the buoyancy of the eddy and Newton's laws, and the mean free path, $\ell$, referred to as the mixing length, is the free parameter of the theory. It is typically taken to be some fraction of the pressure scale height. A number of uncertainties attend this attempt to model convection and an attempt will be made to describe these below.

Some sort of convective motions will occur if a fluid is unstably stratified, that is, if a displaced fluid element finds itself subjected to a buoyancy force tending to amplify the displacement. Whether a fluid element will be unstable, and if so the mean velocity that it will acquire, will depend on the assumptions made as to how the fluid element is displaced. If it is displaced adiabatically (the typical assumption) and at constant composition, then the resulting convection if it occurs is referred to as Ledoux convection. If it moves adiabatically but maintains the same composition as the background, then the resulting convection is referred to as Schwarzschild convection. If the composition gradient is zero the criterion for the two is the same. As the background composition gradient in a star, when nonzero, almost always goes from heavier to lighter elements as a function of radius ({\it e.g.}, in the wake of a retreating convective region), the composition gradient tends to be stabilizing. Thus Ledoux convection is more restrictive, in the sense that a fluid can be unstable to Schwarzschild convection but stable to Ledoux convection.

Regions unstable to Schwarzschild convection but stable to Ledoux convection can be doubly diffusive unstable \citep{kato_66}, a phenomenon usually referred to as semiconvection, although this term has been used to refer to a multitude of sins. A fluid element perturbed outward under these conditions will find itself hotter than the background and therefore tend to continue the displacement, but will be stabilized by its heavier composition. Thermal diffusion, if faster than compositional diffusion, will tend to thermally equilibrate the fluid element with the background while leaving it with a compositional difference tending to drive it back. What can result is an oscillation of the fluid element with growing amplitude. It is unclear how to mix the material under these conditions. Two dimensional numerical simulations \citep{merryfield_95} suggest a complicated situation. Large-amplitude standing waves which break and mix over a distance of the order of a wavelength will arise if the instability is strongly driven. If the instability is weakly driven short waves arise initially and then organize themselves into longer waves which occasionally overturn and mix, and ultimately come to resemble horizontally propagating solitary waves. It is not clear how to connect the results of these simulations, which were unable to reach steady state, with the statistical steady state that presumably develops over the evolutionary time scales of stars. Extreme assumptions among stellar evolution modelers are that semiconvective mixing is fast and the use the Schwarzschild criterion for convection is therefore appropriate, or that it is slow and the use of the Ledoux criterion is therefore appropriate.

Semiconvection originally referred to another ambiguity that arises when a hydrogen burning core moves outward in mass as its helium content grows \citep{schwarzschild_h58}. This happens in some massive star models as the pressure in the convective core becomes more dominated in time by radiation and convective instability is more easily achieved. A chemical discontinuity arises at the convective core boundary. If electron scattering dominates the opacity, as is the case for massive stars, then the opacity increases across the convective core boundary and a problem arises as to the placement of this boundary. As the boundary is approached from the inside the radiative gradient becomes equal to the adiabatic gradient. But just outside the boundary the opacity increase implies that the radiative gradient must exceed the adiabatic gradient. Hence the boundary should be moved outward. Doing so removes the composition gradient, however, and hence removes the need to move the boundary outward in the first place. This poses a dilemma, and a number of schemes have been proposed for dealing with it. These have been summarized by \citet{stothers_70}.

\begin{figure}[ht]
\centerline{\epsfxsize=4.1in\epsfbox{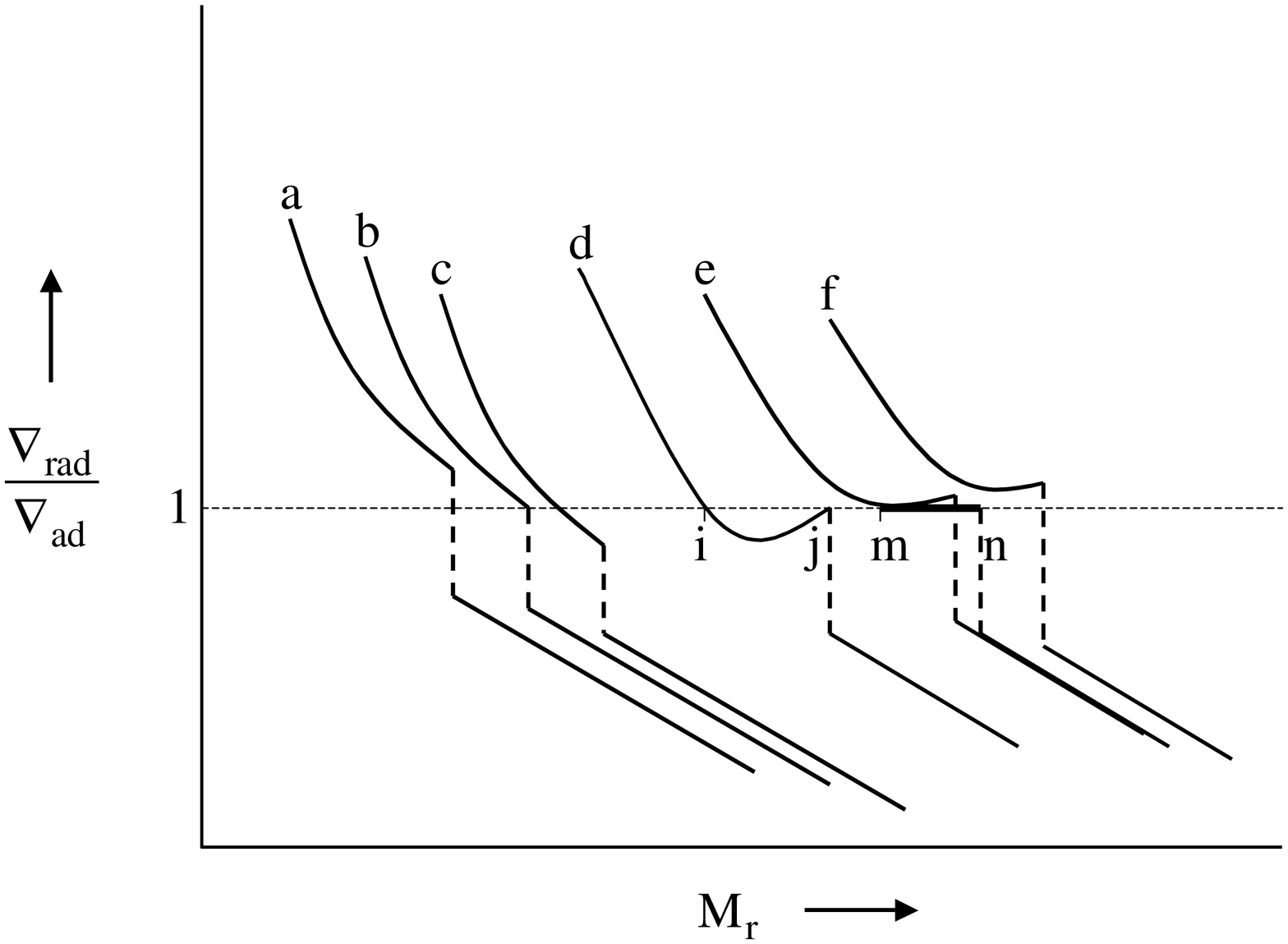}}   
\caption{The curves illustrate profiles of $\nabla_{\rm rad}/\nabla_{\rm ad}$ in the vicinity of the boundary between a convective helium burning core and a radiative hydrogen envelope. Curves a to c illustrate an unambiguous determination of the convective core boundary ({\it e.g.}, curve b). Curves d to f illustrate the ambiguity that arises when a minimum in the opacity occurs within the convective core. (Figure adapted from Fig. 7 of \citep{paczynski_70b}.) \label{Semiconvection3}}
\end{figure}

A related mixing ambiguity, also referred to as semiconvection, can happen in stars with expanding helium burning cores \citep{schwarzschild_h69, paczynski_70b, castellani_gr71a, castellani_gr71b, robertson_71, robertson_f72}. In this case the electron scattering is the same just inside and just outside the core, but the carbon rich  mixture inside the core has a higher free-free opacity. This by itself does not prevent the boundary of the convective core from being located unambiguously, as curves a to c in Fig. \ref{Semiconvection3} illustrate. The convective boundary occurs where the radiative gradient becomes equal to the adiabatic gradient, and curve b has correctly located this boundary. As the helium burning core grows, however, a point is reached where the opacity develops a minimum inside the convective core and then increases outward to the core boundary. In this case the attempt to find the convective core boundary leads to the possibilies illustrated by curves d to f. If curve d is chosen to locate the core boundary, the region between i and j is not convective, contradicting the choice. If curves e or f are chosen, the material at the edge of the core will be unstable to further convection since $\nabla_{\rm rad} > \nabla_{\rm rad}$ there. What is frequently done is to assume that curve e, modified by the horizontal segment connecting points m and n represents the correct choice. This is achieved by assuming that the requisite ``semiconvective'' compositional mixing takes place between points m and n causing $\nabla_{\rm rad} = \nabla_{\rm rad}$ there.

Another problem with the MLT parameterization of convection is overshooting, which refers to the tendency of convective eddies to penetrate the radiative layers surrounding a convective zone and hence induce a mixing of a region larger (in mass) than classically allowed by the strict adoption of the Schwarzschild or Ledoux criterion. This is a consequence of the fact that while the acceleration of the convective motions cease at the boundary of a convective region, there velocities do not. Thus convective overshooting may be present at the border of any convective region, and is not confined to any particular evolutionary phase. The effect of overshooting is to increase the mass in the convective region that is mixed which, in turn, can have a number of consequences for stellar ages, nucleosynthesis, and presupernova structure. Unfortunately, is not a natural outcome of MLT, due to the local nature of the theory \citep{renzini_97}.

The radial extent, $\ell_{\rm OV}$, of the thermal and chemical mixing from the formal convective core boundary is typically parameterized by an ad hoc formula such as $\ell_{\rm OV} = \alpha_{\rm OV} \min (r, H_{\rm p})$, where $H_{\rm p}$ is the pressure scale height, $r$ is the core radius, the distance from the core edge to the surface, or some other such scale that naturally limits the extent of overshooting, and $\alpha_{\rm OV}$ is the parameter of the theory, typically below unity. The extent of convective overshooting will be a function of the P\'{e}clet number \citep[{\it e.g.},][]{zahn_99}, which is the ratio of the convective to the radiative diffusivity. For large P\'{e}clet numbers at the border of a convective region (typical of convective regions well below the stellar surface), the convective eddies exchange little heat with the background and therefore establish a nearly adiabatic gradient beyond the unstable region. They are therefore decelerated by the stable stratification. For small P\'{e}clet numbers, however, radiation diffusion will tend to thermally equilibrate the convective eddies with the background as they penetrate beyond the formal convective boundary which will weaken their deceleration. In this case little heat is transported, but chemicals and momentum can be transported an appreciable distance. (Technically, the former (large P\'{e}clet number) case is referred to as convective penetration, the latter is referred to as convective overshooting \citep{zahn_91}.)

Some observational constraints suggesting a value of $\sim 0.2$ for $\alpha_{\rm OV}$ are provided by  the size of gaps (blue loops) in open star cluster color-magnitude diagrams \citep{maeder_m81, stothers_c91a, stothers_c91b, stothers_91, nordstrom_aa97}, the asteroseismology  of $\eta$ Bootis \citep{dimauro_ckbp03}, accurate stellar dimensions derived from well-detached double-lined eclipsing binaries \citep{gimenez_crj99} (which suggests a somewhat larger value of $\alpha_{\rm OV}$ for massive stars). Beyond this the value of $\alpha_{\rm OV}$ must be inferred from numerical simulations or  guessed at.

A number of numerical simulations investigating the nature of turbulent compressible convection and convective overshooting have been performed. These include two dimensional simulations \citep{hulburt_tm86, hulburt_tm94, dintrans_bns03}, three dimensional simulations  \citep{cattaneo_btmh91, muthsam_gklz95, singh_rc_98, stein_n98, brummell_ct02}, three dimensional simulations with rotation \citep{brummell_ht96, browning_bat04}, three dimensional simulations with ionization \citep{rast_nst93, rast_t93a, rast_t93b}, and two dimensional simulations \citep{bazan_a94, bazan_a98}. and they reveal a rather complicated picture. Depending on the density contrast, upward-moving flows are typically broader and slower moving than downward-moving flows (a trend seen in the neutrino driven convecting regions in post collapse stellar cores). Ionization regions can exagerate this trend. The downward flows traverse multiple scale heights and penetrate the stable layers below by a significant fraction of the local pressure scale height. Because of the low filling factor of the plumes, however, they do not establish an adiabatic gradient there. Convective overshooting can excite gravity waves which leads to further mixing. The use of MLT during shell oxygen burning and later nuclear burning stages is particularly problematic, as nuclear burning timescales at the base of the convecting region and convective timescales across the convective region become comparable. The simulations show inhomogeneities in the composition and strong fluctuations in space and time unlike the smooth, steady flow presupposed by one-dimensional stellar evolutionary calculations with MLT.

In conclusion we observe that MLT is a phenomenological parameterization of convection which is applied to a variety of convective phenomena in a physically motivated but crude way. Different prescriptions for MLT convection can lead to substantial differences in the interior structure of massive stars in their late evolutionary phases. We note just one example. The nonrotating models computed by \citet{hirschi_mm04}, who used the Schwarzschild criterion for convection with overshooting, and with convective diffusion beyond He burning, develop considerably larger Si core masses than the models computed by \citet{rausherhhw02}, who used the Ledoux criterion for convection with semiconvection and without overshooting, and convective diffusion beyond He burning. 

\subsection{ $^{12}$C($\alpha,\gamma$)$^{16}$O Reaction Rate and Convection }

The structure and explosive yields of massive stars depends on the mass fraction, $X_{\rm C}$, of $^{12}$C left after He burning, and this depends both on the combined effects of the  $^{12}$C($\alpha,\gamma$)$^{16}$O reaction rate and the treatment of overshooting and semiconvection which governs the growth of the helium burning core \citep{weaverw93, thielemannnh96, imbrianilgtsc01}. The triple-$\alpha$ reaction and the $^{12}$C($\alpha,\gamma$)$^{16}$O reaction compete with each other, and the ratio of the two rates determines directly the ratio of $^{12}$C and $^{16}$O produced during core helium burning. The mixing of fresh He fuel into the He-burning core at late stages and high temperatures, when a core without growth by semiconvection and overshooting of convection eddies would have already ceased to process any He fuel, probes the $^{12}$C rate at higher temperature with the effect of turning much of the remaining $^{12}$C into $^{16}$O. However, despite years of effort, the $^{12}$C and $^{16}$O cross section is still unknown to within a factor of a few \citep{buchmann_abhl96, angulo_etc99}. Furthermore, as discussed above the treatment of convection in stellar evolutionary codes is by means of MLT, which is rudimentary and phenomenological, and cannot address the question of convective overshooting. 

The implication of the uncertainty in the value of $X_{\rm C}$ left after helium burning is that its value affects  the later structure of the star principally through its effect during the time interval that elapses between the end of the central C burning and the beginning of the central Ne burning. During this time the CO core experiences a phase of gravitational contraction which  is partially alleviated by the formation of one (or more) convective C shell episodes. These convective episodes stop for a while the outwardly advancing C-burning front while the reservior of fuel contained in the convective shell is consumed. During this time the C-burning front remains essentially fixed in mass and slows down the contraction of the region above the front. A larger value of $X_{\rm C}$ after core carbon burning allows a more effective support of the layers above the C-burning front during C-shell burning and hence the formation of a less steep mass-radius relation. These differences in the mass-radius relations that form before the Ne ignition remain through later core contractions until the final explosion.

\begin{figure}[ht]
\centerline{\epsfxsize=4.1in\epsfbox{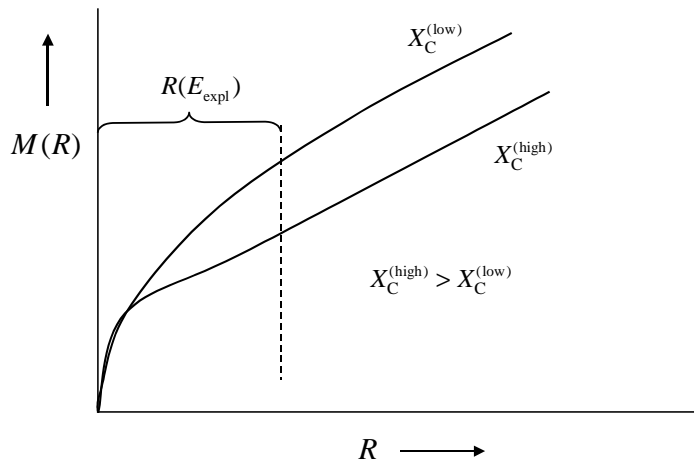}}
\caption{The enclosed mass as a function of radius, $R$, of two initially identical main sequence stars at the onset of core collapse. The $X_{\rm C}^{\rm low}$ and $X_{\rm C}^{\rm high}$ curves represent the case of a low and a high mass fraction of $^{12}$C after core He-burning. \label{XC12}}
\end{figure}

The situation is shown schematically in Fig. \ref{XC12}, which is adapted from Fig. (12) of \citet{imbrianilgtsc01}. The lines denoted by $X_{\rm C}^{\rm high}$ and $X_{\rm C}^{\rm low}$ represent the mass-radius relations for a star at the onset of core collapse having a high and a low value of $X_{\rm C}$, respectively, after core He-burning. Once the explosion commences and the shock wave moves outward, it is radiation dominated, gains or loses only a small fraction of its energy to the matter, and therefore expands essentially adiabatically \citep{weaverw80}. The temperature behind the shock is therefore a function only of its radius and the explosion energy. At the same time, the matter which is subject to complete silicon burning, incomplete silicon burning, or oxygen burning, and whose final composition therefore depends only on its initial proton fraction, $Y_{\rm e}$, is determined only by the peak post shock temperatures, and therefore by the geometrical distance of the matter from the core center. Because of the mass-radius relation (Fig.  \ref{XC12}), the mass of material undergoing incomplete explosive Si burning and explosive O burning, which produce the bulk of the elements from $^{28}$Si to $^{55}$Mn, is greater for small $X_{\rm C}$. On the other hand, the lighter elements from $^{20}$Ne to $^{27}$Al are produced in the C convective shell and partial destroyed by the shock, and their production therefore scales with $X_{\rm C}$. Ignoring the subtleties of many of the production chains, it is seen that a large $X_{\rm C}$ favors the production of elements at the lighter end of the $^{20}$Ne to $^{55}$Mn range, while the opposite is true of a small $X_{\rm C}$. A dramatic illustration of this trend is shown in Fig. 4 of \citet{weaverw93}, who tried to put limits on  the $^{12}$C($\alpha,\gamma$)$^{16}$O reaction rate by requiring that the final explosive yields to have a scaled solar relative abundance. (\citet{arnett_71} made an analogous attempt to fix the $^{12}$C($\alpha,\gamma$)$^{16}$O reaction rate by using the observed  $^{12}$C to $^{16}$O ratio.) It must be remembered that these attempts to fix the $^{12}$C($\alpha,\gamma$)$^{16}$O reaction rate by using the results of stellar evolutionary calculations actually fix a combination of this rate and the particular MLT scheme used.

Further problems with  MLT convection for the evolution of massive stars, in brief, are the fact that it fails to deal with the interaction between convection and rotation, and the generation and transport of magnetic fields \citep{zahn_99}, and convective nuclear burning \citep{bazan_a94, bazan_a98, asida_a00}. Attempts have been made to overcome the limitations of MLT by a first principles approach ({\it i.e.}, direct numerical solutions of the fundamental equations) \citep[{\it e.g.},][]{stein_n98, singh_rc_98, deupree_98, asida_a00}, or by more sophisticated convection models \citep[{\it e.g.},][]{canuto_m91, canuto_m92}, but these have not made there way into evolutionary calculations of massive stars to core collapse.

\subsection{ Weak Interactions }

Weak interactions affect Y$_{\rm e}$, the proton fraction, and therfore play an important role in determining both the presupernova stellar structure and the nucleosynthesis. They affect the structure because, at all times, the pressure is mostly due to electrons - at first, nonrelativistic and nondegenerate, but later neither. They affect the nucleosynthesis because the synthesis of all nuclei except those with equal numbers of neutrons and protons is sensitive to Y$_{\rm e}$.

The weak interaction rates after oxygen burning are particularly difficult to calculate as a large number of excited states with uncertain properties become populated so that their decay must be dealt with statistically. Early attempts in this direction were made by \citet{hansen_68, mazurek_tc74, takahashi_yk73}. However, it was \citep{fuller_fn80, fuller_fn82a, fuller_fn82b, fuller_82, fuller_fn85}  who recognized the key role played by the Gamow-Tellar resonance and noted that measured decay rates exploited only a small fraction of the available strength. More recently new shell-model calculations of the distribution of Gamow-Tellar strength have resulted in an improved---and often reduced---estimate of its strength \citep{martinez_pinedo_l99, langanke_p00, martinez_pinedo_ld00, langanke_m03}. Inclusion in presupernova evolutionary models of these new rates for electron capture and beta dacay \citep{hegerwml01} lead to slightly higher central proton fractions and smaller outer core entropies at the time of core collapse, leading to slightly smaller iron core masses. Incorporation of the new rates in core collapse simulations \citep{langanke_msdhmmljr03, hix_mmlsldm03} lead to an increased importance of nuclear vs free proton electron capture and reduced initial mass behind the shock with lower densities, proton fractions, and entropies. However, the reduced electron capture in the outer layers slows their collapse and allows the shock to reach a slightly larger maximum radius. The collapsing core encounters a range of large and neutron rich nuclei whose beta strengths have not yet been calculated in detail, underscoring the need for more work in this area.

\subsection{ Rotation }
\label{sec: Rotation}

Massive stars are observed to be rapid rotators, with equatorial velocities spanning the range $100 - 400$ km $s^{-1}$ \citep{fukuda_82, halbedel_96, penny_96, howarth_shgp97}. As a consequence, a number of instabilities leading to composition mixing and angular momentum transport are predicted to occur within these stars as they evolve, leading to differences in the structure of supernova progenitors. Furthermore, the rotation rate of progenitor cores may play a role in the supernova mechanism and is dependent on the degree to which angular momentum transport has occurred during the course of prior evolution.

A number of observations point to the operation of rotationally induced mixing processes in massive stars. The ratio B/R, the number of blue to red supergiants, increases with the metalicity, Z,  \citep[{\it e.g.},][]{langer_m95, maeder_m00a}, and this cannot be accounted for by mass loss or convection. For a number of reasons \citep{maeder_m01} rotation favors the development of the red supergiant structure. The increase of the B/R ratio with Z results from the increase of the mass loss rate with Z, and with it the loss of angular momentum of the star, rapidly reducing its rotation rate during the MS phase and thus reducing its propensity to become a red supergiant during later phases. Rotational mixing in the radiative envelopes of massive stars will modify their surface abundances. One would naively expect a depletion of an initial surface abundance of fragile nuclei, such as  $^{3}$He, $^{6}$Li, $^{7}$Li, $^{9}$Be,  $^{10}$B, and $^{11}$B, mixed down and destroyed by proton capture at higher interior temperatures. At the same time, hydrogen burning in massive stars is governed by the CNO cycle, and this has the effect of converting most of the initial $^{12}$C and $^{16}$O into $^{14}$N. Rotational mixing to the surface of material in which the CNO cycle was operative should be manifested as a depletion of $^{12}$C and $^{16}$O and an enhancement of $^{14}$N. These effects have been observed. For example, \citet{proffitt_q01, venn_bllllk02} have observed boron depletions in B type stars in OB associations, consistent with the predictions of \citet{fliegner_lv96} and the rotating models of \citet{heger_l00}. Some non-supergiant B stars show a moderate increase in N abundance \citep{gies_l92, lennon_dmpm96}.

Ideally, the evolution of rotating stars should be calculated multi-dimensionally, with the composition and angular momentum transport arising directly from the calculation itself. This program cannot be carried out with current computer resources. Rather, the equations of stellar structure are kept one-dimensional. Initially this was accomplished by replacing the usual spherical coordinates by new coordinates characterizing the equipotentials (which have cylindrical symmetry) \citep{kippenhahn_ht70}. More recently, this is accomplished by making the assumption \citep{zahn_92} of highly anisotropic turbulence in radiative layers. In particular,  turbulence generated by, say, shear in the presence of differential rotation is much stronger in the direction perpendicular to gravity ("horizontal direction") than in the vertical direction, the latter being suppressed by the stable vertical stratification. If true, the strong horizontal turbulence makes the angular velocity $\Omega$ and the composition nearly constant on isobaric surfaces, rather than cylinders, giving rise to a ``shellular'' rotation law.  In this case, the motion is not cylindrical. Nevertheless, a consistent 1-D scheme has been formulated \citep{meynetm97, meynet_m00}.

The critical assumption of highly anisotropic turbulence in radiative stellar zones has indirect observational support, both in the fact that turbulent motions caused by shear stresses in the Earth's atmosphere are highly anisotropic in those regions where the stratification is stable, and in the study of the solar tachocline \citep{spiegel_z92}. (The tachocline is the transition zone between the rigid rotation in the radiative interior and the external convective zone, where rotation varies with latitude.) If the horizontal turbulence is intense, then the tachocline is very thin, and the  latter is supported by helioseismological observations.

Keeping the equations of stellar structure one-dimensional allows stellar evolutionary calculations to be performed, but requires that various instabilities leading to angular momentum transport and the mixing of chemical elements, which play a major role in massive star evolution, be parameterized. Since the diffusion and advection of composition and angular momentum operate on $\Omega$ and $(r \sin \theta)^{2} \Omega$, respectively, their vertical transport rates are different, being much smaller for the composition. Gradients in composition ($\mu$-gradients) that develop during the evolution of the star tend to reduce the vertical transport rates, so the effect of these $\mu$-gradient effects must either also be parameterized \citep{hegerlw00} or attempt to incorporate them more consistently in the instability and mixing algorithms \citep{maeder_z98}.

Rotation in convective zones is relatively easy to handle until oxygen shell and particularly silicon burning. Chemical homogeneity can be assumed and, if the  high viscosity associated with turbulence tends to solid-body rotation, then rigid body rotation can also be assumed. An alternative \citep{endals76} is that convection preserves the angular momentum of the convective elements leading to equalization of the specific angular momentum. Supporting the tendency towards rigid body rotation over alternatives, however, is the observation that the solar convection zone deviates from solid body rotation by less than 5\% \citep{antia_bc98}. Complications in handling convective zones begin with oxygen shell burning. The times scales for convective mixing, nuclear burning and angular momentum transport become similar, and the feasibility of constructing self-consistent models with one-dimensional evolution equations becomes highly suspect.


At convective boundaries and in radiative zones a number of instabilities can lead to significant transport of composition and angular momentum \citep[{\it e.g.},][and many others]{endals78, hegerlw00, meynet_m00, maeder_m00a}. These include the Eddington-Sweet circulation \citep{vonzeipel_24, eddington_25, vogt_25} (a circulation that arises because a component of the radiation stress is directed along equipotential surfaces), shear instabilities \citep{spiegel_z70, zahn_74} (dynamical: arising when the free energy in differentially rotating layers exceeds the work against restoring forces required to adiabatically overturn the fluid; secular: as above but allowing thermal diffusion in the overturning fluid to remove a stabilizing temperature gradient), the Solberg-H{\o}iland instability \citep{tassoul_00} (analogous to the Ledoux criterion but including the angular momentum gradients), and the Goldreich-Schubert-Fricke instability \citep{goldreich_s67, fricke_68} ((1) a secular analogue to the Solberg-H{\o}iland stability criterion, and (2) a criterion for the generation of meridional flows for nonconservative rotation profiles).

During the main-sequence evolution of rapidly rotating massive stars, angular momentum transport in the outer radiative regions, principally by the Eddington-Sweet circulation, quickly establishes a steady state in which the diffusion of angular momentum is balanced by advection of angular momentum due to circulation \citep{zahn_92, urpin_ss96, talon_zmm97}. Neglecting angular momentum loss at the surface, this leads to a differential rotation in which the angular velocity at the convective core boundary is about a factor of 1.15 that at the surface.

Concerning the late evolutionary stages, which are of most interest to supernova modelers, there have been two recent stellar evolutionary calculations of rotating stars that have been carried out to these stages (without magnetic fields). These are by \citet{hegerlw00} (HLW), who evolve to core collapse, and \citet{hirschi_mm04} (HMM), who evolve through central oxygen or silicon burning. The two groups employed different numerical methods of incorporating the effects of rotation, and different parameterizations of convection.

HMM find that for stars with zero age main-sequence masses (M$_{\rm MS}$) $<$ 30 M$_{\odot}$ rotation tends to increase the mass of the iron core prior to collapse. This trend is confirmed by HLW who confine their study to M$_{\rm MS} < 25$ M$_{\odot}$. The larger iron cores in the rotating models result from rotational mixing in prior evolutionary phases, mainly the H-burning phase where the mixing and nuclear timescales are comparable. The larger He cores resulting from the rotational mixing during H-burning cause them to have lower densities and higher temperature during He-core-burning, and this leads to lower C to O ratios at core He exhaustion. HLW, who parameterize the inhibiting effect of $\mu$-gradients on mixing, find on varying this parameter that the final iron core masses are a sensitive functions of this parameter. The more efficient the rotational mixing (or less strong are the inhibiting effects of the $\mu$-gradients), the greater the core masses. (This underscores the fact that the precollapse structure of rotating stars is highly dependent on approximate numerical treatments of complicated physics.) The larger precollapse iron core models resulting from rotation imply, of course, a smaller initial M$_{\rm MS}$ that will lead to core collapse. HMM obtain significantly greater core masses then HLW, due to their different treatments of rotational mixing and convection, again underscoring the effect of different approximations. Both groups find that only convective process are rapid enough to notably redistribute angular momentum after core helium burning The effect is to leave each convective region with a constant angular velocity. The distribution of angular momenta in the final models is therefore characterized by rounded saw-tooth patterns, each saw tooth being the constant angular velocity imprint of a convective zone. For M$_{\rm MS}$ $>$ 30 M$_{\odot}$, HMM find that rotationally enhanced mass loss causes the star to enter the Wolf-Rayet phase earlier and to therefore spend more time undergoing heavy mass loss. This erodes the star and results in smaller cores at the pre-supernova stage.

The two groups find that the angular momentum of precollapse cores tend to converge to a value that would imply a rotation rate upon collapse to a neutron star of about 1 ms or less, which is near breakup for even the slowest rotators. The corresponding angular momentum is about 100 times the angular momentum of the fastest rotating pulsars observed. The implication of this in unclear, as it is not yet established how rapidly newly formed pulsars rotate. For example, the fastest rotating young pulsar, PSR J053726910 \citep{marshall_gzmw98}, has a period of 16 ms. But with its estimated age of $\sim 5 \times 10^{3}$ yr an extrapolation to an initial rotation rate of $\sim 1$ ms, while not the only possibility,  is not unreasonable. The Crab (and by implication others), on the other  hand, may never have rotated near breakup \citep{trimble_r70} unless this rotational energy ($\sim 3\times 10^{52}$ ergs) were radiated as gravitational waves. Otherwise this energy would surely have been seen in the optical and manifested now in the expansion velocity of the nebula. 

\subsection{ Rotation and Mass Loss }

Mass loss from the stellar surface (stellar winds) significantly affects the evolution of massive stars, particularly the Upper MS stars {\citep{chiosi_m86, maeder_m87} where an appreciable percentage of the mass of these star can be lost during their evolution. Type 1b and 1c supernovae probably arise from stars having suffered extensive mass loss. Mass loss also affects rotating stars, as these winds can transport large amounts of angular momentum out of the star. This is particularly true for equatorial mass loss by anisotropic stellar winds \citep{maeder_99}. However, the uncertainties in these mass-loss rates, particularly for red supergiants \citep{lamers_c99} are considerable due both to the uncertain physics involved and the uncertainties in the observational data and their interpretation. (For example, mass loss rates from hot O and B stars have rencently been revised, generally downward \citep{nugis_l00, vink_kl00, vink_kl01, bouret_lh04}, owing to a improved treatments of clumping and multiple scatering.  

Stellar evolutionary calculations of nonrotating stars with mass find that stars with solar metalicity initially more massive than $\sim 30 - 35$ M$_{\odot}$ converge to a hydrogen free star of roughly 5 M$_{\odot}$ \citep{maeder_90, schaller_smm92,woosley_lw93, meynet_mssc94}. This assumes the loss of the hydrogen envelope either during the main sequence phase (as luminous blue variables) or as red supergiants), and a mass loss rate for hydrogenless Wolf-Rayet stars being given by a positive power of the remaining mass \citep{langer_89a, langer_89b}. The latter causes the convergence in mass. However, while the masses of these stars may converge, the thermal and chemical structures of these stars retain some memory of their former masses \citep{woosley_lw93}. For example, a helium core undergoing carbon burning and trimming down from a larger mass will have a higher temperature and lower density than a constant mass helium core of the same final mass. It will therefore burn more of its helium to oxygen than its constant mass counterpart. Thus, despite similar final iron core masses, stars that have trimmed down from larger initial masses will have larger carbon-oxygen mantles with larger oxygen to carbon ratios and shallower density gradients in the outer regions than stars of the same final mass that have trimmed down from smaller initial masses. Nonrotating stars that do not succeed in loosing their hydrogen envelopes during hydrogen and core helium burning have internal structures that are little affected by the mass loss, as the interior evolution is largely decoupled from the surface. However, these stars may give rise to different supernova types (Type IIL versus Type IIP, for example) depending on the remaining mass of the hydrogen envelope.

Rotating stars undergoing mass loss can lose considerable angular momentum during evolution \citep{packet_vlsg80, heger_l98}. The mass loss itself is affected by rotation through centrifugal forces, nonradial forces, and gravitational darkening (von Zeipel's theorem). The mass loss rate is increased by rotation, but it is now appreciated that it could be either oblately ({\it i.e.}, predominantly equatorial) or prolately ({\it i.e.}, predominantly polar) asymmetrical \citep{owocki_cg98, petrenz_p00}, with much less angular momentum being lost by the star for a given amount of mass lost in the latter case.

\subsection{ Rotation and Magnetic Fields }

\begin{figure}[ht]
\centerline{\epsfxsize=2.1in\epsfbox{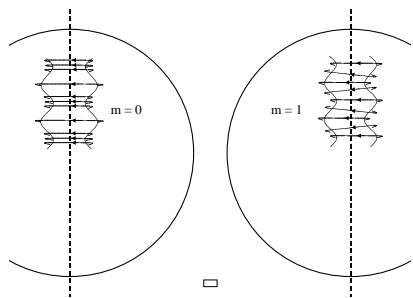}}   
\caption{Illustration of $m$ = 0 and $m$ = 1 instabilities near the axis of a star, where $m$ is the azimuthal mode of number of the perturbation. In stably stratified (radiative) zones, the $m$ = 1 instability is more likely to occur as it involves displacements close to equipotential surfaces.  (Figure adapted from Fig. 1 of \citep{taylor_73}.) \label{B_perturb}}
\end{figure}

The growth of magnetic fields and their influence on the evolution of rotating stars has recently begun to be considered in stellar evolutionary calculations of stars to core collapse. The prime  importance of magnetic fields here is in their potential for transferring angular momentum. As discussed in the previous Section, without magnetic fields the various hydrodynamic instabilities leading to angular momentum transport (as presently understood) are two weak to prevent the cores of supernova progenitors from ending up with about 100 times the angular momentum of the shortest period young pulsar. While this large angular momentum might be appropriate for the collapser model of gamma-ray bursts \citep{macfadyen_wh01}, it is likely too high to account for pulsar rotation rates at birth. Magnetic effects might provide additional angular momentum transport during stellar evolution. A possible case in point is the Sun, whose near-uniform rotation of the radiative core and the small difference in rotation rate between the core and the convective envelope, the latter being continuously spun down by the solar wind torque, has been established through helioseismology \citep{corbard_bmpst97, schou_etal98}. On the other hand, it has been known for a long timed that hydrodynamic instabilities alone are incapable of accounting for this \citep[{\it e.g.},][]{spruit_kr83}.

In convective zones, angular momentum transport by turbulent viscosity is very efficient. Where the transport of angular momentum by magnetic fields may be critical is in radiative zones. \citet{spruit_99, spruit_02} has summarized much of the literature pertaining to this issue and described how a magnetic dynamo based on the an instability studied by \citet{taylor_73} and others and tapping the free energy available from differential rotation could operate in stably stratified zones. The basic picture is that differential rotation will wind up an initially weak field producing a predominantly toroidal (azimuthal) field. This field is subject to a number of instabilities, the m = 1 Taylor instability, exhibited in Fig. \ref{B_perturb}, being likely the most relevant. The poloidal component generated by this instability will be stretched into a strong toroidal component which will in turn be subjected to instabilities, forming a dynamo. An estimate of the equilibrium strength of the field generated by this ``Taylor-Spruit'' dynamo is made by equating the growth timescale to the attenuation timescale by mangetic diffusivity. The radial component over the largest unstable lengths is chosen to evaluate these timescales as this determines the maximum saturation field. The result is obtained for the two limiting cases in which the stability is provided either by a thermal or $\mu$ gradient.

Comparisons of estimates of the angular momentum transport by the magnetic fields generated by the Taylor-Spruit dynamo with the angular momentum transport by hydrodynamic instabilities indicate that the former could be of major importance \citep{maeder_m03}. A recent stellar evolutionary calculation \citep{heger_wls04} incorporating the estimates of magnetic torques produced by the Taylor-Spruit magnetic fields show a reduction by a factor of 10 in the final iron core angular momentum. It must be observed, however, that angular momentum transport by magnetic fields has the potential of being completely ineffective in a star or so effective as to lead to near uniform rotation throughout the entire star with the result of a much too slow rotation of the remnant \citep{spruitp98}. Thus, while the above results are encouraging, the fact that magnetic fields have an almost ``just so'' effectiveness must regarded as extremely preliminary.

\subsection{ Global Asymmetries }

A potentially important effect in the late evolutionary stages of massive stars, which would require a multidimensional evolutionary code to follow, is the generation of overstable g-modes driven by shell nuclear burning \citep{goldreich_ls96}. The idea is that nuclear burning rates in silicon and oxygen burning shells are extremely temperature sensitive. Consider an $\ell = 1$ perturbation of the shell to the right, for example. The left-hand side of the shell will then be compressed and heated. Nuclear burning in the compressed region will be greatly enhanced and will generate a large local overpressure which will  push the shell back to the left. If overstable, this  g-mode will oscillate with growing amplitude with the very interesting possibility of generating a significant global asymmetry of the stellar core just prior to collapse.

\subsection{ Conclusions }

The evolutionary calculations of massive stars to core collapse are extremely difficult, involving a variety of physics on multiple length and time scales and in multiple dimensions. Teams are incorporating more realistic physics into the stellar codes, but are still constrained to model inherently multidimensional phenomena in one dimension. This review has attempted to make supernova modelers aware of the many approximations and parameterizations that are perforce made in the course of evolving a star from the main-sequence to core collapse.

\newpage

\bibliography{BibTeX_list}

\end{document}